\documentclass[12pt]{article}
\usepackage[top=26truemm,bottom=28truemm,left=23truemm,right=23truemm]{geometry}

\usepackage{indentfirst}

\usepackage{mathrsfs,amsmath,amsfonts,amssymb,mathtools,physics,latexsym,bm}
\usepackage{array,empheq}
\usepackage{authblk}
\usepackage{appendix}
\usepackage[pdftex]{graphicx}
\usepackage{fancybox,ascmac,here}
\usepackage{multirow}
\usepackage[dvipsnames]{xcolor}
\usepackage[colorlinks,linkcolor=blue,anchorcolor=violet,citecolor=red]{hyperref}
\usepackage{float}
\usepackage{comment,cancel}
\usepackage[normalem]{ulem}
\usepackage{caption,subcaption}
\usepackage{url}

\usepackage[numbers]{natbib}

\numberwithin{equation}{section}

\begin{document}
\begin{titlepage}
\unitlength = 1mm
\begin{flushright}
KOBE-COSMO-25-12
\end{flushright}

\vskip 1cm
\begin{center}

{ \Large \textbf{ Toward graviton detection \\
via 
photon-graviton quantum state conversion
}}
\vspace{1.8cm}
\\
Taiki Ikeda$^*$, Youka Kaku$^{\flat}$, Sugumi Kanno$^*$, and  Jiro Soda$^{\flat}$
\vspace{1cm}

{\it $^*$ Department of Physics, Kyushu University, Fukuoka 819-0395, Japan \\
\it $^\flat$ Department of Physics, Kobe University, Kobe 657-8501, Japan
}

\vskip 4.0cm

{\large Abstract}\\
\end{center}

A magnetic field enables the interconversion of photons and gravitons, yet the process is usually analysed only at the level of classical wave equations. We revisit photon-graviton conversion in a quantum field theoretic framework, allowing us to track the evolution of arbitrary quantum states. Treating the photons as squeezed coherent states and the gravitons as the squeezed vacuum expected for primordial gravitational waves, we derive analytic expressions for the conversion probability and show that it can be significantly enhanced compared to the conventional estimate. We further demonstrate that the conversion both swaps preexisting entanglement and generates genuinely new entanglement between the electromagnetic and gravitational sectors, which is impossible in any classical description. Detecting such nonclassical correlations would constitute compelling evidence for the quantization of gravity and offers a novel pathway toward graviton detection.

\vspace{1.0cm}
\end{titlepage}

\hrule height 0.075mm depth 0.075mm width 165mm
\tableofcontents
\vspace{1.0cm}
\hrule height 0.075mm depth 0.075mm width 165mm

\section{Introduction}

One of the long-standing questions in fundamental physics is whether the graviton can be detected~\cite{Dyson:2013hbl}.
Dyson argued that the enormous number of occupations of typical gravitational wave backgrounds renders individual gravitons effectively unobservable. Remarkably, this same observation suggests two complementary search strategies. First, one can exploit large occupation numbers and look for gravitons through the quantum noise they induce in sensitive detectors~
\cite{Parikh:2020nrd,Kanno:2020usf,Parikh:2020kfh,Parikh:2020fhy,Kanno:2021gpt}.  Second, one can target high-frequency gravitational waves, where the occupation number is much smaller. The latter idea underlies axion-magnon experiments repurposed for high-frequency gravitational-wave detection~\cite{Ito:2019wcb,Ito:2020wxi,Ito:2022rxn}, and
has inspired subsequent photon-based proposals
~\cite{Ejlli:2019bqj,Berlin:2021txa,Domcke:2022rgu,Kanno:2023whr}. In the noise-based approach, quantum entanglement plays a central role; in the frequency-based approach, graviton–magnon or graviton–photon conversion is the key mechanism. The central aim of this
work is to combine these two perspectives. For concreteness, we focus on photon–graviton conversion, but the formalism also extends naturally to graviton–magnon and graviton–phonon conversion~\cite{Tobar:2024bjr}.

Recall that the linear gravitational perturbation $h_{ij}$ couples to the electromagnetic field via the minimal-coupling term
\begin{align}
     \frac{1}{M_{\rm pl}} h_{ij} 
     \left( E^i E^j -B^i B^j \right) ~,
\end{align}
where  $E^i$ and  $B^i$ are the electric and magnetic fields and $M_{\rm pl}$ is the Planck mass.
In the presence of a static magnetic field, this interaction enables photon–graviton conversion (and its inverse process)~\cite{Gertsenshtein:1962,Raffelt:1987im,Chen:1994ch,Cillis:1996qy}. 
For a photon of arbitrary frequency propagating through a homogeneous field, the conversion probability is (see, e.g.,~\cite{Masaki:2018eut})
\begin{align}
   P(\gamma \rightarrow g ) 
   = \sin^2 \Delta_{\rm osc}L,
\end{align}
with propagation length $L$ and
\begin{align}
    \Delta_{\rm osc}  =\frac{B}{\sqrt{2}M_{\rm pl}}     ~.
\end{align}
Here we have ignored plasma effects and the electron one-loop correction.
For intuition, 
consider $B=10\,{\rm T}$ and $L=10^4\,{\rm km}$, for which the conversion probability becomes extremely small as shown below:
\begin{align}
    P(\gamma \rightarrow g ) 
  \simeq  \left(\frac{BL}{\sqrt{2}M_{\rm pl}} \right)^2\,\simeq\, 10^{-20} 
\label{convention}  ~.
\end{align}
Accordingly, enhancing the conversion probability and devising detection strategies 
are essential
toward a practical graviton search.

In conventional treatments, photon–graviton conversion is analyzed solely at the level of classical wave equations, so any nonclassical features of the photon field are neglected. In contrast, modern quantum-sensing techniques routinely exploit nonclassical light to enhance detector sensitivity~\cite{RevModPhys.89.035002}. 
Motivated by this, we reexamine photon–graviton conversion from a quantum field theoretic perspective, incorporating the quantum nature of both photons and gravitons~\cite{Carney:2023nzz,Kharzeev:2025lyu,Ikeda:2025qac}. 
Our formalism accommodates squeezed coherent states for photons and two-mode squeezed states for gravitons. We show that, in a background of primordial gravitational waves, the conversion probability can be enhanced by a factor of order $10^4$. Building on this result, we then investigate how the conversion process reshapes the entanglement structure of a state. In particular, we demonstrate that photon–graviton conversion can both swap existing entanglement and generate new entanglement effects that are impossible in purely classical dynamics. Observing such behavior would therefore constitute strong evidence that gravity is quantized. In this way, photon–graviton conversion offers a novel and potentially powerful avenue for graviton searches.

The organization of the paper is as follows. In section 2, we introduce the photon-graviton system with a constant magnetic field background. In section 3, we reproduce the conventional photon-graviton conversion probability in a quantum field theoretical manner. In section 4, we apply the formalism to the case of nonclassical photons such as the squeezed state. Remarkably, we find a significant enhancement of the conversion probability in the background of primordial gravitational waves generated during inflation. We also investigate how the entanglement structure changes after conversion.
This might be indirect evidence for the existence of gravitons if detected. 
The final section is devoted to the conclusion.

\section{Graviton-photon system}

In this section, we consider the interaction between photons and gravitons in the presence of a constant magnetic field, assuming Minkowski spacetime background. We then examine linear perturbations of both electromagnetic and gravitational fields. Strictly speaking, a constant background magnetic field curves spacetime. However, throughout this paper, we neglect such backreaction effects.

The total action for the gravitational field $g_{\mu\nu}$ and the electromagnetic field $F_{\mu\nu}$ is given by
\begin{align}
    S=S_g+S_A=\frac{M_{\rm pl}^2}{2}\int d^4x \sqrt{-g}
    \,
    R-\frac{1}{4}\int d^4x \sqrt{-g}\,
    F_{\mu\nu}
    F^{\mu\nu}
    \label{eq:original action}
    ~,
\end{align}
where $M_{\rm pl}=1/\sqrt{8\pi G}$ is the Planck mass. The electromagnetic field strength tensor is defined in terms of the gauge field $A_\mu$ as $F_{\mu\nu}=\partial_\mu A_\nu-\partial_\nu A_\mu$.
We now consider linear perturbations of the field based on this action.

The perturbation of the gravitational field around Minkowski spacetime is given by
\begin{align}
    ds^2
    =g_{\mu\nu}dx^\mu dx^\nu
    =- dt^2+\left(\delta_{ij}+h_{ij}\right)dx^idx^j\,,
\end{align}
where $t$ is the time coordinate and $x^i\,(i=1,2,3)$ are the spatial coordinates.
The metric perturbations $h_{ij}$ satisfy the transverse-traceless (TT) conditions:  $\partial_j h^{ij}=0$ and $h^i{}_i=0$. 
In what follows, spatial indices are raised and lowered using the flat background metric $\delta_{ij}$, rather than  $\delta_{ij}+h_{ij}$. 
 
Perturbation of the electromagnetic field in the presence of a constant background magnetic field
is introduced as follows:
\begin{align}
    &A_\mu =\bar A_\mu +\delta A_\mu \,, \\
    &\bar F_{ij}=\partial_i \bar A_j-\partial_j \bar A_i
    = \epsilon_{ijk}B^k \,,
    \qquad
    \bar F_{\mu 0}=0
    ~.
\end{align}

Here, $\bar A_\mu$ denotes the background gauge field, and accordingly, $B^k$ is the constant background magnetic field. The perturbative electromagnetic field satisfies the Coulomb gauge conditions: $\delta A_0=0$ and $\partial_i\delta A^i=0$.

The Einstein-Hilbert action expanded to second order in $h_{ij}$
is given by
\begin{align}
    \delta S_g =
    \frac{M_{\rm pl}^2}{8}
    \int d^4 x \left[
    \dot h_{ij}\,\dot h^{ij} -\partial_k h_{ij}\,\partial^k h^{ij}
    \right]
    ~,
\end{align}
where a dot denotes a derivative with respect to time. Again, note that the raising and lowering of spatial indices in the above expression are performed using $\delta_{ij}$, so that higher-order contributions
in $h_{ij}$
are omitted.
From the above expression, we can see that $\psi_{ij}=M_{\rm pl}\, h_{ij}/2$ corresponds to the canonically normalized field. Introducing its canonical momentum $\pi^{ij}=\dot{\psi}^{ij}$,
we proceed with canonical quantization by imposing the following commutation relation:
\begin{align}
    [\psi_{ij} (t , x^i) , \pi^{k\ell} (t,  y^i)]
    =\frac{i}{2}\left( \bar{P}_i{}^k \bar{P}_j{}^\ell
   + \bar{P}_i{}^\ell \bar{P}_j{}^k - \bar{P}_{ij} \bar{P}^{k\ell}\, \right)
   \delta^{(3)} ( x^i - y^i)  ~,
   \label{eq:commut_h}
\end{align}
where the transverse projection tensor is defined as
\begin{align}
    \bar{P}_{ij}  = \delta_{ij} -\frac{\partial_i \partial_j}{\nabla^2}  ~.
\end{align}

The action for the photon field, expanded to second order in perturbations 
$\delta A_i$, is given by
\begin{align}
    \delta S_A=\frac{1}{2}\int d^4x\left[{\delta\dot{A}}^i \, \delta\dot{A}_i-
    \partial^i\delta A^k\,\partial_i\delta A_k
    \right]
    ~.
    \label{action:A}
\end{align}
The canonical momentum conjugate to $\delta A_i$ is $\pi^i = \delta\dot{A}^i$.
The canonical commutation relation is
\begin{align}
    [\delta A_{i} (t ,  x^i) , \pi^{k} (t, y^i)]
    = i \bar{P}_i{}^k\,\delta (x^i - y^i)  ~.
    \label{eq:commut_A}
\end{align}

The action for the interaction between gravitons and photons, expanded to second order in the perturbations $h_{ij}$ and $\delta A^i$, takes the following form:

\begin{align}
    \delta S_{\rm I}=
    \varepsilon_{i\ell m}B_m \int d^4x \left[h^{ij}\left(
    \partial_j \delta A_\ell
    -\partial_\ell \delta A_j
    \right)
    \right]~.
    \label{action:I}
\end{align}
Note that $B_m=\varepsilon_{mj\ell}\,\partial_j \bar A_\ell$ represents a constant background magnetic field.

We now work in Fourier space and use the wavevector ${\bm k}$.
We first introduce the polarization vectors $e^P_i(\bm{k})$ for the gauge field, where $P=+$ or $\times$ labels the two polarization modes.
They satisfy the orthonormality condition and the completeness relation:
\begin{align}
    &e^{Pi}(\bm{k}) e_i^{Q}(\bm{k})=\delta^{PQ} ~,\\
    \label{eq:orthornormality_ei}
    &\sum_P e^P_i(\bm{k}) e^{Pj}(\bm{k})=P_i{}^j ~,
\end{align}
where we defined the transverse projection tensor in Fourie space
\begin{align}
    P_{ij}  = \delta_{ij} -\frac{k_i k_j}{k^2}  ~.
\end{align}
Note that the completeness relation is consistent with the Coulomb gauge condition $\partial_i\delta A^i=0$. In particular, it implies $P_i{}^j e^P_j=e^P_i$, confirming that the polarization vectors are transverse to the wavevector $\bm{k}$. 
Thus, the polarization vectors $e_i^{+}, e_i^{\times}$ and the normalized wavevector $k^i/|\bm{k}|$ form an orthonormal basis.

The polarization tensors for $h_{ij}$ can be expressed in terms of the  polarization vectors  $e_i^{P}(\bm{k})$ as
\begin{align}
    &e^+_{ij}(\bm{k})
    =\frac{1}{\sqrt{2}} \left(
    e^+_i(\bm{k}) e^+_j(\bm{k})-e^\times_i(\bm{k}) e^\times_j(\bm{k})
    \right)~,\\
    &e^\times_{ij}(\bm{k})
    =\frac{1}{\sqrt{2}}\left(
    e^+_i(\bm{k}) e^\times_j(\bm{k})+e^\times_i(\bm{k}) e^+_j(\bm{k})
    \right)\,.
\end{align}
The polarization tensors also satisfy the orthonormality condition and the completeness relation:
\begin{align}
    &e^{Pij}(\bm{k})\,e^Q_{ij}(\bm{k})=\delta^{PQ}~,\\
    &\sum_P e_{ij}^P (\bm{k})\,e^{P k\ell }(\bm{k}) =
   \frac{1}{2}\left( P_i{}^k P_j{}^\ell
   + P_i{}^\ell P_j{}^k - P_{ij} P^{k\ell} \,\right) ~.
\end{align}
This completeness relation is consistent with the tranverse-traceless gauge. 
In what follows, we adopt the convention:
\begin{align}
    e^+_i(-\bm{k})=e^+_i(\bm{k})~,\qquad
    e^\times_i(-\bm{k})=-e^\times_i(\bm{k})~.
    \label{eq:right_handed}
\end{align}

The field operator $h_{ij}(t,\bm{x})$, which satisfies the canonical commutation relation, 
can be expanded in terms of the annihilation and creation operators as
\begin{align}
    h_{ij}(t,\bm{x})=\frac{2}{M_{\rm pl}} \sum_{P=+,\times}\frac{1}{(2\pi)^{3/2}} \int d^3 k\,
    \left[ h^{P}_{\bm k}(t)\, e^{P}_{ij}(\bm{k}) a_P ({\bm k}) + h^{P*}_{-\bm k}(t)\, e^{P}_{ij}(-\bm{k}) a_P^\dagger  ({-\bm k})\right] \,e^{i\bm{k}\cdot\bm{x}}~,
\end{align}
where we define the mode function as
\begin{align}
    h^P_{\bm k}(t)  
    = \frac{1}{\sqrt{2\omega_g}} ~e^{-i\omega_g t}  ~,
    \label{eq:h_modefunc}
\end{align}
with the dispersion relation $\omega_g = |{\bm k}|$.
From the canonical commutation relation of the field operator in Eq.~\eqref{eq:commut_h}, we obtain the following commutation relation for the annihilation and creation operators
\begin{align}
    [a_P(\bm{k}),a^\dag_Q(\bm{k}')]=\delta_{PQ}\,\delta^{(3)}(\bm{k}-\bm{k}')
    ~,
\end{align}
where $P,Q=+,\times$.
Similarly, the field operator $\delta A_i(t,\bm{x})$, which satisfies the canonical commutation relation, can be expanded as
\begin{align}
    \delta A_i(t,\bm{x})=\sum_{P=+,\times} \frac{1}{(2\pi)^{3/2}}
    \int d^3 k\,\left[ A^{P}_{\bm k}(t)\,e^{P}_i(\bm{k}) \, b_P ({\bm k}) + A^{P*}_{-\bm k}(t)\,e^{P}_i(-\bm{k}) \, b_P^\dagger ({-\bm k}) \right]\,e^{i\bm{k}\cdot\bm{x}}\,.
\end{align}
For the electromagnetic fields, the mode function is given by 
\begin{align}
    A^{P}_{\bm k}(t)  
    = \frac{1}{\sqrt{2\omega_\gamma}} e^{-i\omega_\gamma t}  ~,
    \label{eq:A_modefunc}
\end{align}
where the dispersion relation for photons is $\omega_\gamma =|{\bm k}|$. Again, using the canonical commutation relation in Eq.~\eqref{eq:commut_A}, we obtain the following
\begin{align}
    [b_P(\bm{k}),b^\dag_Q(\bm{k}')]=\delta_{PQ}\,\delta^{(3)}(\bm{k}-\bm{k}')\,.
\end{align}

\begin{figure}[htbp]
    \centering
    \includegraphics[width=0.4\linewidth]{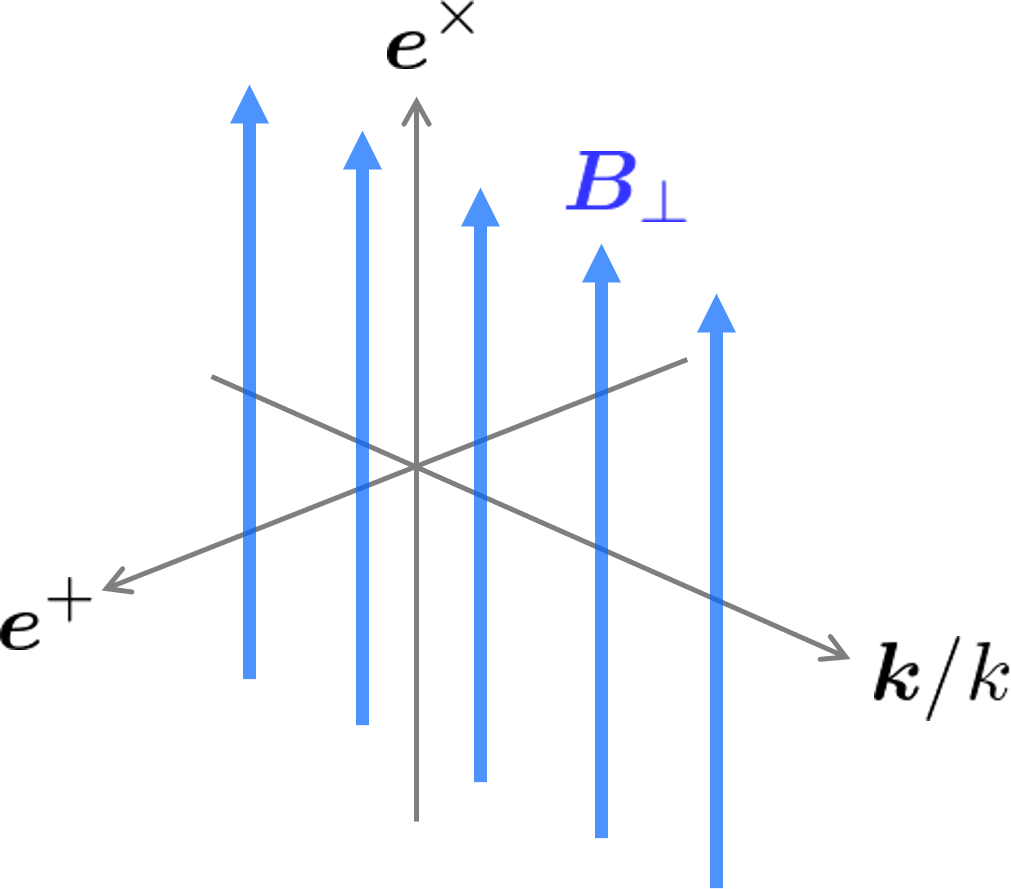}
    \caption{Three vectors $\bm{k}/k,~\bm{e}^\times$ and $\bm{e}^+$ form a right-handed coordinate system.}
    \label{fig:right_handed}
\end{figure}

We can now rewrite the interaction action Eq.~\eqref{action:I}  as
\begin{align}
    \delta S_I &= -\frac{2}{M_{\rm pl}}\varepsilon_{i\ell m}\,B^m\,\sum_{P,Q}\int d^3 k\,dt
    \left[
    h_{\bm k}^P \, e^{Pij}(\bm k) \,a_P(\bm k)  
    +h_{\bm k}^{P*} \, e^{Pij}(-\bm k) \, a^\dagger_P(-\bm k)  \right]\nonumber\\
    &\qquad \times 
    \left[ A_{\bm k}^{Q} \, e^Q_{j}(-\bm k) \, b_{Q}(-\bm k)  
    +A_{\bm k}^{Q*} \, e^Q_{j}(\bm k) \, b^\dagger_{Q}(\bm k) 
    \right]\left(-ik^\ell\right)
    ~,
    \label{eq:SI_2}
\end{align}
where $k=|\bm k|$. 
We decompose the background magnetic field as follows:
\begin{align}
    \bm{B} = \bm{B}_\parallel + \bm{B}_\perp,
    \qquad
    B_{\parallel \,i} = \frac{k_ik_j}{k^2} B^j,
    \qquad
    B_{\perp \, i} = P_{ij} B^j\,.
\end{align}
The component $\bm{B}_\parallel$ does not contribute to the graviton-photon conversion, as it vanishes in the term $\varepsilon_{ilm}B_\parallel^mk^\ell$ in Eq.~\eqref{eq:SI_2}.
Hence, in what follows, we focus only on $\bm{B}_{\perp}$.
Without loss of generality, we assume that $\bm{B}_{\perp}$ points in the $\bm{e}^{ \times}$ direction.
As shown in Fig.~\ref{fig:right_handed}, the unit vector ${\bm k}/k$ and the polarization vectors $\bm{e}^{\times}, \bm{e}^{+}$ form a right-handed coordinate system, corresponding to the $x,~y$ and $z$ directions, respectively.
Since these vectors form an orthonomal basis, we can show that
\begin{align}
    \varepsilon_{ilm}\,B_\perp^m e^{Pij}(\bm k)
   e^Q_{j}(-\bm k) k^l
   \propto \delta^{PQ} ~.
\end{align}
Then, the double sum over $P$ and $Q$ in Eq.~(\ref{eq:SI_2}) reduces to a single sum over $P$. Moreover, using Eqs.~\eqref{eq:h_modefunc},~\eqref{eq:A_modefunc} and the orthonormal condition of the polarization vector given by Eq.~\eqref{eq:orthornormality_ei}, we can further reduce the action as
\begin{align}
    S_I
    &= i\lambda \int d^3k\,dt\,
    \left[ 
    a_+(\bm k) b^\dagger_{+}(\bm k)
    + a_\times(\bm k) b^\dagger_{\times}(\bm k)
    - \{ a^\dagger_+(\bm k) b_{+}(\bm k) + a^\dagger_\times(\bm k) b_{\times}(\bm k) \}
    \right.\nonumber\\
    &\left. 
    +e^{-2ikt} \{a_+(\bm k) b_{+}(-\bm k)-a_\times(\bm k) b_{\times}(-\bm k)
    \} -  e^{2ikt} \{
    a^\dagger_+(\bm k) b^\dagger_{+}(-\bm k) - a^\dagger_\times(\bm k) b^\dagger_{\times}(-\bm k)
    \}
    \right]
    ~.
    \label{eq:SI_3}
\end{align}
Here, we define the coupling between graviton and photon as 
\begin{align}
    \lambda  \equiv\frac{1}{\sqrt{2}M_{\rm pl}}
    \varepsilon^{ilm}\,e^+_i\,\frac{k_l}{k}\,B_{\perp m}
    =  \frac{B_\perp}{\sqrt{2}M_{\rm pl}}   
    ~,
    \label{eq:coupling}
\end{align}
where $B_\perp=|\bm{B}_\perp|$.
Note that $\lambda$ is a pseudoscalar.

\section{Conversion of photon into graviton in quantum field theory}

In this section, we revisit the standard result for photon-graviton conversion probability using a quantum field theoretic approach.

\subsection{Interaction Hamiltonian}

We adopt the interaction picture.  
Using the result from the previous section, the interaction Hamiltonian can be written as
\begin{align}
    H_I
    &= -i\lambda \int d^3k\,
    \left[ 
    a_+(\bm k) b^\dagger_{+}(\bm k)
    + a_\times(\bm k) b^\dagger_{\times}(\bm k)
    - \{ a^\dagger_+(\bm k) b_{+}(\bm k) + a^\dagger_\times(\bm k) b_{\times}(\bm k) \}
    \right.\nonumber\\
    &\left. 
    +e^{-2ikt} \{a_+(\bm k) b_{+}(-\bm k)-a_\times(\bm k) b_{\times}(-\bm k)
    \} -  e^{2ikt} \{
    a^\dagger_+(\bm k) b^\dagger_{+}(-\bm k) - a^\dagger_\times(\bm k) b^\dagger_{\times}(-\bm k)
    \}
    \right]
    ~.
    \label{eq:HI}
\end{align}
This interaction generates mixing between photons polarized along $e_i^{\times}$ and the graviton in the $\times$ mode, as well as between the photon polarized along $e_i^{+}$ and the graviton in the $+$ mode. 

The time evolution operator for the system is calculated from the interaction Hamiltonian in the interaction picture as
\begin{align}
    U(t,0) = T \exp \left[ -i\int^t_0 H_I (t') ~dt'\right]\,,
\end{align}
where $T$ denotes the time-ordering. Since the time dependence only appears in the c-number coefficients of $H_I$, the time-ordering operator $T$ can be omitted as follows:
\begin{align}
    U(t,0) = \exp[-iQ]
    ~,
    \label{eq:U}
\end{align}
where
\begin{align}
    Q 
    &= \int^t_0 H_I(t') ~dt' \nonumber\\
    &= -i\lambda \sum_{P=+,\times} \int d^3k\,
    \left[ 
    t\{
    a_P(\bm k) \, b^\dagger_P(\bm k) - a^\dagger_P(\bm k) \, b_P(\bm k)
    \}
    \right.\nonumber\\
    &\left.\hspace{40mm}
    +f^P_k(t) \, a_P(\bm k) \, b_P(-\bm k) - f^{P*}_k(t) \, a_P^\dagger(\bm k) \, b_P^\dagger(-\bm k)
    \right]
    ~,
\end{align}
and 
\begin{align}
    f^+_k(t) = \frac{ \sin kt}{k}~e^{-ikt} ~,
    \qquad
    f^\times_k(t) = -\frac{ \sin kt}{k}~e^{-ikt} ~.
\end{align}
Thus, the $+$ and $\times$ sectors are decoupled.

\subsection{Photon-graviton conversion probability }

We consider the probability that a single photon ($\gamma$) is converted into a single graviton ($g$): 
\begin{eqnarray}
    |0\rangle_g |1_{\bm{k},P} \rangle_\gamma \rightarrow |1_{\bm{k},P}\rangle_g |0 \rangle_\gamma
    ~.
\end{eqnarray}    
Here, $|n_{\bm{k},P}\rangle_g$ and $|n_{\bm{k},P}\rangle_\gamma$ are Fock states with $n$ gravitons or photons of momentum ${\bm k}$ and polarization $P$, defined by the creation and annihilation operators $a_P^\dagger(\bm{k})$, $a_P(\bm{k})$ and $b_P^\dagger(\bm{k})$, $b_P(\bm{k})$, respectively.
Using the time evolution operator in Eq.~\eqref{eq:U}, the transition probability is given by
\begin{align}
    P \left( \gamma \rightarrow g \right)
    &=\left| 
    {}_g\langle 1_{\bm{k},P}| \, {}_\gamma\langle0| \, e^{-iQ} \, |0\rangle_g \, |1_{\bm{k},P} \rangle_\gamma
    \right|^2  \nonumber\\
    & = \left| {}_g\langle 0| \, {}_\gamma\langle 0| \, a_P(\bm k) \, e^{-iQ} \, b_P^\dagger(\bm k) \, |0 \rangle_g |0\rangle_\gamma  \right|^2 \nonumber\\
    & = \left| \lambda \, t \right|^2 
    ~.
\end{align}
Using Eq.~\eqref{eq:coupling}, the transition probability is given by
\begin{align}
    P \left( \gamma \rightarrow g \right)
    =  \left(\frac{B_\perp L}{\sqrt{2} M_{\rm pl}} \right)^2 
    ~,
\end{align}
where we have identified the interaction time $t$ with the interaction length $L$.

In this section, we have calculated the conversion probability using a field theoretical approach and reproduced the conventional result~(\ref{convention}). We are now in a position to take into account the quantum state of the photon in the conversion process.

\section{Entanglement generation via photon-graviton conversion}

In this section, we consider photon-graviton conversion in nontrivial quantum states. Experimentally, it is possible to prepare a squeezed coherent state of the photon field. On the other hand, assuming the inflationary scenario, the graviton field is expected to be in a squeezed vacuum state. In this setting, we show that the conversion probability can be significantly enhanced.

\subsection{Various quantum states of photons and gravitons}
In this subsection, we discuss photon quantum states with minimum uncertainty. The analysis below applies to gravitons as well.
We begin by considering the coherent state, which can be generated by a classical current. 
The coherent state is constructed using the displacement operator
\begin{align}
    D_{\bm{k},P}(\beta)  
    = \exp\left[ \beta_{\bm{k},P} \,b_P^\dagger(\bm k) 
    - \beta^*_{\bm{k},P} b_P(\bm k) \right]
\end{align}
and defined as
\begin{align}
    |\beta_{\bm{k},P} \rangle_\gamma = D_{\bm{k},P}(\beta) \, |0 \rangle_\gamma  ~. 
\end{align}
This is the eigenstate of the annihilation operator:
\begin{align}
    b_P(\bm k) \, |\beta_{\bm{k},P} \rangle_\gamma
    = \beta_{\bm{k},P}  |\beta_{\bm{k},P} \rangle_\gamma \,, 
\end{align}
where the coherent parameter $\beta_{\bm{k},P}$ generally depends on the wave vector $\bm{k}$ and the polarization mode $P$.
The coherent state is the quantum state closest to a classical state.

Similarly, the squeezed state is constructed using the squeezing operator
\begin{align}
    S_{\bm{k},P}(\zeta) 
    = \exp\left[ -\frac{1}{2} \left( 
    \zeta^*_{\bm{k},P} \, b_P(\bm k)^2 - \zeta_{\bm{k},P} \,b_P^\dagger(\bm k)^2
    \right)\right] ~, 
\end{align}
with $\zeta_{\bm{k},P} = r_{\bm{k},P} \,\exp\left(i\varphi_{\bm{k},P}\right)$ called the squeezing parameter. The squeezed state is then given by
\begin{align}
    |\zeta_{\bm{k},P} \rangle_\gamma = S_{\bm{k},P}(\zeta) |0 \rangle_\gamma  
    ~. 
\end{align}
Here, $r_{\bm{k},P} \in \mathbb{R}$ controls the degree of squeezing (i.e., how much quantum uncertainty is reduced in one quadrature and increased in the other), while $\varphi_{\bm{k},P}$ determines the orientation of the squeezing ellipse in phase space. In other words, the squeezing parameter describes how the quantum fluctuations of the vacuum are deformed. 
It is useful to note how the squeezing operator transforms the annihilation operator: 
\begin{align}
   S_{\bm{k},P}^{\dagger}(\zeta) b_P(\bm k) S_{\bm{k},P}(\zeta)
   = b_P(\bm k) \cosh r_{\bm{k},P} + b_P^\dagger(\bm k) \,e^{i\varphi_{\bm{k},P}} \sinh r_{\bm{k},P}
   ~.
   \label{eq:squeeze_b}
\end{align}
This relation shows that squeezing mixes the annihilation and creation operators, reflecting the quantum correlations introduced by the squeezing operation.

The squeezed coherent state is constructed as 
\begin{align}
   |\zeta_{\bm{k},P} , \, \beta_{\bm{k},P} \rangle_\gamma =S_{\bm{k},P}(\zeta) \, D_{\bm{k},P}(\beta)\,|0\rangle_\gamma 
   ~.
\end{align}
This state incorporates both squeezing and displacement, and represents a more general quantum state of light than a purely coherent or squeezed state.

\subsection{Conversion in squeezed coherent states of photons}

Let us assume that the initial photon state is a single photon added to a squeezed coherent state.
Since this state is not normalized, we need incorporate the normalization constant $A_\gamma$ as
\begin{eqnarray}
A_\gamma \, b_P^\dagger(\bm{k}) \, |\zeta_{\bm{k},P}, \, \beta_{\bm{k},P}\rangle_\gamma\,. 
\end{eqnarray}
The normalization constant $A_\gamma$ is evaluated as
\begin{align}
    A_\gamma =\Bigl(
    \cosh^2 r_{\bm{k},P}+|\beta_{\bm{k},P}|^2\left(
    \cosh 2r_{\bm{k},P} + \cos\left[2\arg\beta_{\bm{k},P}-\varphi_{\bm{k},P}\right] \, \sinh 2r_{\bm{k},P}
    \right)
    \Bigr)^{-1/2} \ .
\end{align}

Now, let us consider the transition probability for a single photon converting into a single graviton, starting from a photon state with one particle added to a squeezed coherent state:
\begin{align}
A_\gamma \,|0\rangle_g\,b_P^\dagger(\bm{k})|\zeta_{\bm{k},P}, \beta_{\bm{k},P}\rangle_\gamma
\longrightarrow
|1_{\bm{k},P}\rangle_g \ |\zeta_{\bm{k},P}, \beta_{\bm{k},P}\rangle_\gamma ~.
\end{align}
Using the time evolution operator in Eq.~\eqref{eq:U},  the transition probability is given by 
\begin{align}
    P\left(\gamma \rightarrow g\right)&=
    A_\gamma ^2
    \left|
    {}_g\langle 1_{\bm{k},P}| \, {}_\gamma\langle \zeta_{\bm{k},P}, \, \beta_{\bm{k},P}|
    \, e^{-i Q} \, b_P^\dagger(\bm{k})\,
    |0\rangle_g \,
    |\zeta_{\bm{k},P}, \, \beta_{\bm{k},P} \rangle_\gamma 
    \right|^2 \nonumber\\
    &=
    A_\gamma^2
    \left|
    {}_g\langle 0| \, {}_\gamma\langle \beta_{\bm{k},P}|
    \,a_P(\bm{k})\, S_{\bm{k},P}^\dagger(\zeta) \, e^{-i W_P} \, b_P^\dagger(\bm{k})\, S_{\bm{k},P}(\zeta) \,
    |0\rangle_g \,
    |\beta_{\bm{k},P} \rangle_\gamma 
    \right|^2
    ~,
\end{align}
where
\begin{align}
    W_P 
    = i\lambda \int d^3k ~ a^\dagger_P(\bm k)
    \left[ 
    t \, b_P(\bm k)
    + f^{P*}_k(t) \, b_P^\dagger(-\bm k)
    \right]
    ~.
\end{align}
Using the relation 
\begin{align}
S_{\bm{k},P}^\dagger(\zeta) \, e^{-i W_P} \,S_{\bm{k},P}(\zeta)= \exp[-i\,S_{\bm{k},P}^\dagger(\zeta) \,W_P \,S_{\bm{k},P}(\zeta)]
\end{align}
and Eq.~\eqref{eq:squeeze_b}, we obtain
\begin{align}
    P\left(\gamma \rightarrow g\right)
    &=
    A_\gamma^2
    \Big|\,{}_\gamma\langle \beta_{\bm{k},P}|
    \Bigl(b_P(\bm k) \cosh r_{\bm{k},P} + \beta_{\bm{k},P}^* \, e^{i\varphi_{\bm{k},P}} \sinh r_{\bm{k},P}\Bigr)\notag\\
    &\hspace{40mm}\times
    \left(b_P^\dagger(\bm k) \cosh r + \beta_{\bm{k},P} \, e^{-i\varphi_{\bm{k},P}} \sinh r_{\bm{k},P}\right)
    |\beta_{\bm{k},P} \rangle_\gamma \,
    \Big|^2
    \\
    &=\left(\frac{B_\perp L}{\sqrt{2}M_{\rm pl}}\right)^2
    \biggl(\cosh^2 r_{\bm{k},P}+|\beta_{\bm{k},P}|^2\left(
    \cosh 2r_{\bm{k},P} + \cos\left[2\arg\beta_{\bm{k},P}-\varphi_{\bm{k},P}\right] \, \sinh 2r_{\bm{k},P}
    \right)
    \biggr)
    ~.
    \nonumber
\end{align}
Thus, photon displacement enhances the transition rate by a factor of $|\beta_{\bm{k},P}|^2$, while the photon squeezing provides an additional enhancement proportional to $e^{2r_{\bm{k},P}}$.
This result can be interpreted as the photon-graviton conversion occurring in a background of squeezed coherent photons.
Conventionally, photon squeezing is often expressed in decibels (dB), defined as $10\log_{10} e^{2r_{\bm{k},P}}$.
Currently, it is possible to generate about $8$ dB of squeezing in the labotatory~\cite{Kashiwazaki:2023pxy}, which corresponds to $e^{2r_{\bm{k},P}}\sim 6.3$.
For $15$ dB squeezed light, as realized in \cite{Vahlbruch:2016sqz}, the enhancement factor reaches approximately $e^{2r_{\bm{k},P}}\sim 40$.

\subsection{Conversion in the primordial gravitational wave background}

During inflation, quantum fluctuations of the gravitational field are stretched and amplified, leading to the generation of primordial gravitational waves~\cite{Grishchuk:1989ss,Grishchuk:1990bj}. These waves evolve into two-mode squeezed quantum states, characterized by strong correlations between $\bm{k}$ and $- \bm{k}$ modes. This squeezing affects the properties of the gravitational wave background and can significantly influence processes such as photon-graviton conversion, potentially enhancing conversion probabilities due to the altered quantum environment. Hence, the graviton state can be described as a two-mode squeezed vacuum state:
\begin{align}
    |\xi_{\bm{k},P} \rangle_g = S_{\bm{k},P}(\xi) |0 \rangle_g  
    ~. 
\end{align}
where the two-mode squeezing operator is defined by
\begin{align}
    S_{\bm{k},P}(\xi) 
    = \exp\left[\, - 
    \xi^*_{\bm{k},P} \, a_P(\bm k)a_P(-\bm k) +\xi_{\bm{k},P} \,a_P^\dagger(\bm k)a_P^\dagger(-\bm k)
    \,\right] ~,\quad  
    \xi_{\bm{k},P} = z_{\bm{k},P} \,\exp\left(i\chi_{\bm{k},P}\right)
    ~.
\end{align}
Here, $\xi_{\bm{k},P}$ is the squeezing parameter for the graviton mode, with $z_{\bm{k},P}$ representing the squeezing amplitude and $\chi_{\bm{k},P}$ the squeezing angle.
In the conventional inflationary scenario, the squeezing amplitude behaves as
\begin{align}
    \sinh 2z_{\bm{k},P} \simeq \cosh 2z_{\bm{k},P} \simeq (k_c/k)^4
    ~,
\end{align}
where $k_c = 2\pi f_c$ is comoving cutoff scale, and $f_c$ is the corresponding cutoff frequency
for primordial gravitational waves (see, e.g., \cite{Kanno:2018cuk}). Observational constraints from the cosmic microwave background (CMB) place an upper bound on this cutoff frequency, typically $f_c \leq 10^9$
Hz. 

In the primordial gravitational wave background, the conversion of photon into a single graviton, starting from a squeezed coherent state:
\begin{align}
A_\gamma \,|\xi_{\bm{k},P} \rangle_g \, b_P^\dagger(\bm{k}) \, |\zeta_{\bm{k},P}, \, \beta_{\bm{k},P}\rangle_\gamma \to A_g \, a_P^\dagger(\bm{k})\, |\xi_{\bm{k},P}\rangle_g\, |\zeta_{\bm{k},P}, \, \beta_{\bm{k},P}\rangle_\gamma
\end{align}
can be calculated as
\begin{align}
    &P\left(\gamma \rightarrow g\right)\nonumber\\
    &=A_\gamma^2\,  A_g^2\left|
    {}_g\langle \xi_{\bm{k},P}| \, {}_\gamma\langle \zeta_{\bm{k},P}, \, \beta_{\bm{k},P}|
    a_P(\bm{k})\, e^{-i Q} \, b_P^\dagger(\bm{k})\,
    |\xi_{\bm{k},P}\rangle_g \,
    |\zeta_{\bm{k},P}, \, \beta_{\bm{k},P} \rangle_\gamma 
    \right|^2 \nonumber\\
    &=A_\gamma^2 \, A_g^2\left|
    {}_g\langle 0| \, {}_\gamma\langle \beta_{\bm{k},P}| S_{\bm{k},P}^\dagger(\xi)a_P(\bm{k})\, S_{\bm{k},P}^\dagger(\zeta) \, e^{-i W_P} \,S_{\bm{k},P}(\xi)\, b_P^\dagger(\bm{k})\, S_{\bm{k},P}(\zeta) \,
    |0\rangle_g \,
    |\beta_{\bm{k},P} \rangle_\gamma 
    \right|^2
      \nonumber \\
    &=\left(\frac{B_\perp L}{\sqrt{2}M_{\rm pl}}\right)^2 \cosh^2 z_{\bm{k},P}
    \left[
    \cosh^2 r_{\bm{k},P}+|\beta_{\bm{k},P}|^2\left(
    \cosh 2r_{\bm{k},P} + \cos\left[2\arg\beta_{\bm{k},P}-\varphi_{\bm{k},P}\right] \, \sinh 2r_{\bm{k},P}
    \right)
    \right]
    ~,
\end{align}
where we have denoted the normalization constant for the graviton as $A_g =(\cosh z_{\bm{k},P})^{-1}$.
Thus, there is an additional enhancement factor $\cosh^2 z_{\bm{k},P}$.
For example, for photons with frequency around $100$ MHz, this corresponds to an enhancement of approximately $10^4$. This extra enhancement factor $\cosh^2 z_{\bm{k},P}$ arises from the squeezing of the graviton state generated during inflation. Squeezing effectively increases the occupation number of gravitons, so the graviton background behaves like a quantum state with more ``particles" than the vacuum. This boosts the probability of photon-graviton conversion compared to the case with no squeezing. In other words, the primordial gravitational wave background is not a simple vacuum state but a two-mode squeezed state with quantum correlations that amplify the conversion process. The stronger the squeezing parameter $z_{\bm{k},P}$, the larger the factor $\cosh^2 z_{\bm{k},P}$, and hence the greater the enhancement of the conversion probability.

The extra boost from gravition squeezing adds to our earlier findings that photon squeezing and displacement also raise the conversion rate.
Altogether, the total transition probability reflects the combined quantum nature of both photons and gravitons in this environment.

In summary, when the photons are prepared in squeezed coherent states and the gravitons reside in two-mode squeezed vacuum states—as expected for laboratory photons and for primordial gravitational waves from inflation—the probability of photon-to-graviton conversion is significantly enhanced. This highlights the importance of considering realistic quantum states beyond vacuum when analyzing such conversion processes.

\subsection{Entanglement generation through photon-graviton conversion}

Photon–graviton conversion is inherently quantum and therefore alters a system’s entanglement structure.
For notational simplicity, we omit the squeezing and coherent parameters in what follows.

Consider an initial state in which two photon modes, ${\bm k}_1$ and ${\bm k}_2$, are entangled, while the corresponding graviton mode ${\bm k}_1$ is in the vacuum:
\begin{align}
    |\psi \rangle =   \frac{1}{\sqrt{2}}\Bigl(\, |1_{\bm{k_1}} \rangle_{\gamma }\otimes|0_{\bm{k_2}} \rangle_{\gamma}  + |0_{\bm{k_1}} \rangle_{\gamma} \otimes|1_{\bm{k_2}} \rangle_{\gamma} \,\Bigr) \otimes |0_{\bm{k_1}} \rangle_{g}  
    ~.
\end{align}

When the conversion operator acts on the ${\rm k}_1$ photon–graviton pair, it can flip the component
$|1_{\bm{k_1}}\rangle_{\gamma}|0_{\bm{k_2}} \rangle_{\gamma}|0_{\bm{k_1}}\rangle_{g}$ into $|0_{\bm{k_1}}\rangle_{\gamma} |0_{\bm{k_2}} \rangle_{\gamma}|1_{\bm{k_1}}\rangle_{g}$, while leaving the orthogonal component $|0_{\bm{k_1}}\rangle_{\gamma}|1_{\bm{k_2}} \rangle_{\gamma} |0_{\bm{k_1}}\rangle_{g}$ unchanged. The state then becomes

\begin{align}
    |\psi \rangle \to  |0_{\bm{k_1}} \rangle_{\gamma }\otimes\frac{1}{\sqrt{2}}
    \Bigl(\, |0_{\bm{k_2}} \rangle_{\gamma} \otimes |1_{\bm{k_1}} \rangle_{g} +|1_{\bm{k_2}} \rangle_{\gamma} \otimes |0_{\bm{k_1}} \rangle_{g} \Bigr) 
    ~,
\end{align}
demonstrating entanglement swapping: the original photon–photon entanglement has been transferred to photon–graviton entanglement.

Alternatively, starting from a separable state in which the three modes factorise:
\begin{align}
    |\psi \rangle =   |0_{\bm{k_1}} \rangle_{\gamma } \otimes\frac{1}{\sqrt{2}}
    \Bigl(\, |0_{\bm{k_2}} \rangle_{\gamma}  + |1_{\bm{k_2}} \rangle_{\gamma} \Bigr) \otimes |1_{\bm{k_1}} \rangle_{g}  
    ~. 
\end{align}
Here the vacuum of the photon mode with wave vector ${\bm k}_1$, the superposition in the photon mode with wave vector ${\bm k}_2$, and the single-graviton state in the graviton mode with wave vector ${\bm k}_1$ are completely independent. When the conversion operator acts on the ${\bm k}_1$ photon–graviton pair 
$|0_{\bm{k_1}}\rangle_{\gamma} |0_{\bm{k_2}} \rangle_{\gamma}|1_{\bm{k_1}}\rangle_{g}$, it flips to $|1_{\bm{k_1}}\rangle_{\gamma} |0_{\bm{k_2}} \rangle_{\gamma}|0_{\bm{k_1}}\rangle_{g}$, while leaving $|0_{\bm{k_1}}\rangle_{\gamma}|1_{\bm{k_2}} \rangle_{\gamma} |1_{\bm{k_1}}\rangle_{g}$ unchanged, The resulting state is
\begin{align}
    |\psi \rangle \to  \frac{1}{\sqrt{2}}\Bigl(|1_{\bm{k_1}} \rangle_{\gamma }\otimes|0_{\bm{k_2}} \rangle_{\gamma} \otimes |0_{\bm{k_1}} \rangle_{g} + |0_{\bm{k_1}} \rangle_{\gamma} \otimes|1_{\bm{k_2}} \rangle_{\gamma} \otimes |1_{\bm{k_1}} \rangle_{g}\Bigr) 
    ~,
\end{align}
which is manifestly entangled: the photon and graviton occupations are now correlated across the three modes, and the state can be no longer be written as a simple product of individual mode states.
Since such entanglement changes cannot occur in classical processes, the observation of these effects can be regarded as strong evidence for the quantum nature of gravitons.
Detecting and quantifying the entanglement generated through photon–graviton conversion is experimentally demanding, but it offers one of the most promising avenues for verifying the quantum nature of gravity.

One strategy is to perform quantum state tomography on the combined photon–graviton system. By measuring correlations between photon states before and after conversion and comparing them with the theoretically predicted entangled states, one can reconstruct the joint density matrix. Standard entanglement measures, such as concurrence or the von Neumann entropy, can then be extracted.
A complementary approach is to test Bell-type inequalities or other entanglement witnesses~\cite{Bell:1964kc,Horodecki:2009zz}. Any violation of classical bounds in photon–graviton correlations would constitute direct evidence of non-classical entanglement.
Because individual gravitons are extraordinarily difficult to detect, indirect signatures may be more practical. For example, one can monitor changes in photon statistics or coherence properties in carefully engineered setups that incorporate strong magnetic fields or tailored quantum-optical states.

Ongoing advances in quantum technologies—such as ultrasensitive interferometers, high-quality squeezed-light sources, and photon-number–resolving detectors—should steadily improve the prospects for observing such subtle quantum-gravitational effects.
Ultimately, demonstrating entanglement involving gravitons would not only confirm their quantum nature but also deepen our understanding of the interplay between quantum mechanics and gravity.

\section{Conclusion}

Progress in graviton detection will likely require tools and concepts from quantum information science.
Motivated by this, we have revisited photon–graviton conversion from a quantum field theoretic perspective and reproduced the standard conversion probability.

Extending the formalism, we evaluated the conversion probability when the photons are prepared in a squeezed coherent state and, separately, when the gravitons themselves are in a two-mode squeezed state as expected for primordial gravitational waves generated during inflation ~\cite{Kanno:2021vwu}. Our treatment generalizes earlier analyses in de Sitter spacetime ~\cite{Kanno:2022ykw,Kanno:2022kve,Kanno:2023fml}. 
We find that a squeezed primordial gravitational wave background amplifies the photon–graviton conversion rate by a factor $\cosh^2{z}$, which can reach ${\cal O}(10^4)$ at frequencies around $100$ MHz. 

As a first step toward experimental verification, we also examined how the conversion process reshapes the entanglement structure of the system.
In particular, photon–graviton conversion can generate entanglement that cannot arise through any classical mechanism; observing such correlations would constitute strong evidence that gravity is quantized.

Several avenues merit further study. A detailed analysis of the dynamics of entanglement in a realistic laboratory setting is needed, along with concrete experimental proposals.
It would also be intriguing to explore analogous processes involving magnons or phonons (graviton–magnon and graviton–phonon conversion).
We leave these extensions to future work.

\section*{Acknowledgments}
T. I. was supported by Research Support Scholarship from Kuroda Scholarship Foundation. Y.K. was supported by Grant-in-Aid for JSPS Fellows. S.~K. was supported by the Japan Society for the Promotion of Science (JSPS) KAKENHI Grant Numbers JP22K03621, JP22H01220, 24K21548 and MEXT KAKENHI Grant-in-Aid for Transformative
Research Areas A “Extreme Universe” No. 24H00967.
J.\ S. was in part supported by JSPS KAKENHI Grant Numbers  JP23K22491, JP24K21548, JP25H02186.

\bibliographystyle{unsrt}
\bibliography{reference}

\begin{thebibliography}{10}

\bibitem{Dyson:2013hbl}
Freeman Dyson.
\newblock {Is a graviton detectable?}
\newblock {\em Int. J. Mod. Phys. A}, 28:1330041, 2013.

\bibitem{Parikh:2020nrd}
Maulik Parikh, Frank Wilczek, and George Zahariade.
\newblock {The Noise of Gravitons}.
\newblock {\em Int. J. Mod. Phys. D}, 29(14):2042001, 2020.
\newblock arXiv:2005.07211 [hep-th].

\bibitem{Kanno:2020usf}
Sugumi Kanno, Jiro Soda, and Junsei Tokuda.
\newblock {Noise and decoherence induced by gravitons}.
\newblock {\em Phys. Rev. D}, 103(4):044017, 2021.
\newblock arXiv:2007.09838 [hep-th].

\bibitem{Parikh:2020kfh}
Maulik Parikh, Frank Wilczek, and George Zahariade.
\newblock {Quantum Mechanics of Gravitational Waves}.
\newblock {\em Phys. Rev. Lett.}, 127(8):081602, 2021.
\newblock arXiv:2010.08205 [hep-th].

\bibitem{Parikh:2020fhy}
Maulik Parikh, Frank Wilczek, and George Zahariade.
\newblock {Signatures of the quantization of gravity at gravitational wave detectors}.
\newblock {\em Phys. Rev. D}, 104(4):046021, 2021.
\newblock arXiv:2010.08208 [hep-th].

\bibitem{Kanno:2021gpt}
Sugumi Kanno, Jiro Soda, and Junsei Tokuda.
\newblock {Indirect detection of gravitons through quantum entanglement}.
\newblock {\em Phys. Rev. D}, 104(8):083516, 2021.
\newblock arXiv:2103.17053 [gr-qc].

\bibitem{Ito:2019wcb}
Asuka Ito, Tomonori Ikeda, Kentaro Miuchi, and Jiro Soda.
\newblock {Probing GHz gravitational waves with graviton\textendash{}magnon resonance}.
\newblock {\em Eur. Phys. J. C}, 80(3):179, 2020.
\newblock arXiv:1903.04843 [gr-qc].

\bibitem{Ito:2020wxi}
Asuka Ito and Jiro Soda.
\newblock {A formalism for magnon gravitational wave detectors}.
\newblock {\em Eur. Phys. J. C}, 80(6):545, 2020.
\newblock arXiv:2004.04646 [gr-qc].

\bibitem{Ito:2022rxn}
Asuka Ito and Jiro Soda.
\newblock {Exploring high-frequency gravitational waves with magnons}.
\newblock {\em Eur. Phys. J. C}, 83(8):766, 2023.
\newblock arXiv:2212.04094 [gr-qc].

\bibitem{Ejlli:2019bqj}
Aldo Ejlli, Damian Ejlli, Adrian~Mike Cruise, Giampaolo Pisano, and Hartmut Grote.
\newblock {Upper limits on the amplitude of ultra-high-frequency gravitational waves from graviton to photon conversion}.
\newblock {\em Eur. Phys. J. C}, 79(12):1032, 2019.
\newblock arXiv:1908.00232 [gr-qc].

\bibitem{Berlin:2021txa}
Asher Berlin, Diego Blas, Raffaele Tito~D'Agnolo, Sebastian A.~R. Ellis, Roni Harnik, Yonatan Kahn, and Jan Sch\"utte-Engel.
\newblock {Detecting high-frequency gravitational waves with microwave cavities}.
\newblock {\em Phys. Rev. D}, 105(11):116011, 2022.
\newblock arXiv:2112.11465 [hep-ph].

\bibitem{Domcke:2022rgu}
Valerie Domcke, Camilo Garcia-Cely, and Nicholas~L. Rodd.
\newblock {Novel Search for High-Frequency Gravitational Waves with Low-Mass Axion Haloscopes}.
\newblock {\em Phys. Rev. Lett.}, 129(4):041101, 2022.
\newblock arXiv:2202.00695 [hep-ph].

\bibitem{Kanno:2023whr}
Sugumi Kanno, Jiro Soda, and Akira Taniguchi.
\newblock {Search for high-frequency gravitational waves with Rydberg atoms}.
\newblock {\em Eur. Phys. J. C}, 85(1):31, 2025.
\newblock arXiv:2311.03890 [gr-qc].

\bibitem{Tobar:2024bjr}
Germain Tobar, Igor Pikovski, and Michael~Edmund Tobar.
\newblock {Detecting kHz gravitons from a neutron star merger with a multi-mode resonant mass detector}.
\newblock {\em Class. Quant. Grav.}, 42(5):055017, 2025.
\newblock arXiv:2406.16898 [astro-ph.IM].

\bibitem{Gertsenshtein:1962}
M.E. Gertsenshtein.
\newblock {Wave Resonance of Light and Gravitational Waves}.
\newblock {\em J.Exp.Theor.Phys.}, 14:84, 1962.

\bibitem{Raffelt:1987im}
Georg Raffelt and Leo Stodolsky.
\newblock {Mixing of the Photon with Low Mass Particles}.
\newblock {\em Phys. Rev. D}, 37:1237, 1988.

\bibitem{Chen:1994ch}
Pisin Chen.
\newblock {Resonant photon - graviton conversion and cosmic microwave background fluctuations}.
\newblock {\em Phys. Rev. Lett.}, 74:634--637, 1995.
\newblock [Erratum: Phys.Rev.Lett. 74, 3091 (1995)].

\bibitem{Cillis:1996qy}
Analia~N. Cillis and Diego~D. Harari.
\newblock {Photon - graviton conversion in a primordial magnetic field and the cosmic microwave background}.
\newblock {\em Phys. Rev. D}, 54:4757--4759, 1996.
\newblock arXiv:astro-ph/9609200].

\bibitem{Masaki:2018eut}
Emi Masaki and Jiro Soda.
\newblock {Conversion of Gravitons into Dark Photons in Cosmological Dark Magnetic Fields}.
\newblock {\em Phys. Rev. D}, 98(2):023540, 2018.
\newblock arXiv:1804.00458 [astro-ph.CO].

\bibitem{RevModPhys.89.035002}
C.~L. Degen, F.~Reinhard, and P.~Cappellaro.
\newblock Quantum sensing.
\newblock {\em Rev. Mod. Phys.}, 89:035002, Jul 2017.

\bibitem{Carney:2023nzz}
Daniel Carney, Valerie Domcke, and Nicholas~L. Rodd.
\newblock {Graviton detection and the quantization of gravity}.
\newblock {\em Phys. Rev. D}, 109(4):044009, 2024.
\newblock arXiv:2308.12988 [hep-th].

\bibitem{Kharzeev:2025lyu}
Dmitri~E. Kharzeev, Azadeh Maleknejad, and Saba Shalamberidze.
\newblock {QuGrav: Bringing gravitational waves to light with Qumodes}.
\newblock 6 2025.
\newblock arXiv:2506.09459 [gr-qc].

\bibitem{Ikeda:2025qac}
Taiki Ikeda, Sugumi Kanno, and Jiro Soda.
\newblock {Enhancing photon-axion conversion probability with squeezed coherent states}.
\newblock 6 2025.
\newblock arXiv:2506.14354 [quant-ph].

\bibitem{Kashiwazaki:2023pxy}
Takahiro Kashiwazaki, Taichi Yamashima, Koji Enbutsu, Takushi Kazama, Asuka Inoue, Kosuke Fukui, Mamoru Endo, Takeshi Umeki, and Akira Furusawa.
\newblock {Over-8-dB squeezed light generation by a broadband waveguide optical parametric amplifier toward fault-tolerant ultra-fast quantum computers}.
\newblock {\em Appl. Phys. Lett.}, 122(23):234003, 2023.
\newblock arXiv:2301.12658 [quant-ph].

\bibitem{Vahlbruch:2016sqz}
Henning Vahlbruch, Moritz Mehmet, Karsten Danzmann, and Roman Schnabel.
\newblock Detection of 15 db squeezed states of light and their application for the absolute calibration of photoelectric quantum efficiency.
\newblock {\em Phys. Rev. Lett.}, 117:110801, Sep 2016.

\bibitem{Grishchuk:1989ss}
L.~P. Grishchuk and Yu.~V. Sidorov.
\newblock {On the Quantum State of Relic Gravitons}.
\newblock {\em Class. Quant. Grav.}, 6:L161--L165, 1989.

\bibitem{Grishchuk:1990bj}
L.~P. Grishchuk and Yu.~V. Sidorov.
\newblock {Squeezed quantum states of relic gravitons and primordial density fluctuations}.
\newblock {\em Phys. Rev. D}, 42:3413--3421, 1990.

\bibitem{Kanno:2018cuk}
Sugumi Kanno and Jiro Soda.
\newblock {Detecting nonclassical primordial gravitational waves with Hanbury-Brown{\textendash}Twiss interferometry}.
\newblock {\em Phys. Rev. D}, 99(8):084010, 2019.

\bibitem{Bell:1964kc}
J.~S. Bell.
\newblock {On the Einstein-Podolsky-Rosen paradox}.
\newblock {\em Physics Physique Fizika}, 1:195--200, 1964.

\bibitem{Horodecki:2009zz}
Ryszard Horodecki, Pawel Horodecki, Michal Horodecki, and Karol Horodecki.
\newblock {Quantum entanglement}.
\newblock {\em Rev. Mod. Phys.}, 81:865--942, 2009.

\bibitem{Kanno:2021vwu}
Sugumi Kanno and Jiro Soda.
\newblock {Squeezed quantum states of graviton and axion in the universe}.
\newblock {\em Int. J. Mod. Phys. D}, 31(13):2250098, 2022.
\newblock arXiv:2112.14496 [gr-qc].

\bibitem{Kanno:2022ykw}
Sugumi Kanno, Jiro Soda, and Kazushige Ueda.
\newblock {Conversion of squeezed gravitons into photons during inflation}.
\newblock {\em Phys. Rev. D}, 106(8):083508, 2022.
\newblock arXiv:2207.05734 [hep-th].

\bibitem{Kanno:2022kve}
Sugumi Kanno, Ann Mukuno, Jiro Soda, and Kazushige Ueda.
\newblock {Impact of quantum entanglement induced by magnetic fields on primordial gravitational waves}.
\newblock {\em Phys. Rev. D}, 107(6):063503, 2023.
\newblock arXiv:2211.05576 [hep-th].

\bibitem{Kanno:2023fml}
Sugumi Kanno, Ann Mukuno, Jiro Soda, and Kazushige Ueda.
\newblock {A peak in the power spectrum of primordial gravitational waves induced by primordial dark magnetic fields}.
\newblock {\em JCAP}, 05:052, 2023.
\newblock arXiv:2301.13540 [hep-th].

\end{thebibliography}

\end{document}